\documentclass[fleqn,12pt,twoside]{article}

\usepackage{amssymb,amsmath}
\usepackage[all]{xy}
\usepackage{bm}
\usepackage{color}

\newcommand{\bes}{\begin{subequations}\bea}
\newcommand{\ees}{\eea\end{subequations}}
\newcommand{\be}{\begin{equation}}
\newcommand{\ee}{\end{equation}}
\newcommand{\bea}{\begin{eqnarray}}
\newcommand{\ba}{\begin{array}}
\newcommand{\eea}{\end{eqnarray}}
\newcommand{\ea}{\end{array}}

\newcommand{\bs}[1]{\boldsymbol{#1}}

\title{Fact-Checking Ziegler's Maximum Entropy Production Principle beyond the Linear Regime and towards Steady States}

\author{Matteo Polettini, University of Luxembourg \footnote{Address:  Campus
Limpertsberg, 162a avenue de la Fa\"iencerie, L-1511 Luxembourg. E-Mail: matteo.polettini@uni.lu.}}

\begin{document}

\maketitle

\begin{abstract}We challenge claims that the principle of maximum entropy production produces physical phenomenological relations between conjugate currents and forces, even beyond the linear regime, and that currents in networks arrange themselves to maximize entropy production as the system approaches the steady state. In particular: (1) we show that Ziegler's principle of thermodynamic orthogonality leads to stringent reciprocal relations for higher order response coefficients, and in the framework of stochastic thermodynamics, we exhibit a simple explicit model that does not satisfy them; (2) on a network, enforcing Kirchhoff's current law, we show that maximization of the entropy production prescribes reciprocal relations between coarse-grained observables, but is not responsible for the onset of the steady state, which is, rather, due to the minimum entropy production principle.
\end{abstract}

\newpage
\section{Introduction}

This paper presents a critical review of several claims regarding Ziegler's principle \cite{ziegler}, the most basic and uncontroversial formulation of a maximum principle for the entropy production. Analogous principles have already earned observational substantiation in climatology \cite{ozawa,kleidon} (yet contrived), and some of them are most probably destined to partake to the establishment of a fundamental paradigm in Nonequilibrium Statistical Mechanics, that of maximum entropy production (MaxEP). However, it is widely recognized that current derivations of such principles have shaky foundations and not clearly delimited domains of application. To avoid the hitches of wishful thinking, it is necessary from time to time to go back to the literature to dot the i's and cross the t's where necessary. ``This is why we can speak clearly and sensibly about making progress [in science]'', to put it with Popper \cite{popper}.

Currently MaxEP is a broad collection of principles that do not necessarily match with each other and do not comply with the physical intuition encoded in the principle of minimum entropy production (MinEP), which is widely agreed to characterize steady states of systems subject to slightly off-equilibrium constraints, again, despite longstanding controversies (see the author's work \cite{myself} for a unified perspective on MinEP, \cite{bruers2} for remarks concerning electric networks and \cite{luo,ma} for proofs in the case of Markovian systems). Different acceptations of MaxEP are:
\begin{itemize}
\item[-] Phenomenological relations maximize the entropy production (Ziegler's principle) \cite{ziegler} (\cite{martyu}[\S 1.2]);
\item[-] The system's state tends to a maximum of the entropy production \cite{looplaw};
\item[-] The probability of paths in state space maximizes entropy production \cite{dewar};
\item[-] Typical paths in state space produce maximum entropy \cite{swenson,zupa};
\item[-] Maximum entropy production is an inferential algorithm, analogous to MAXENT
 \cite{dewar2,kleidon}.
\end{itemize}
There are crucial differences between all these formulations as it comes to the definition of entropy production, to the nature of the variable quantities with respect to which maximization is performed, to the constraints that have to be specified {(for example, flow constraints with respect to potential constraints~\cite{kawazura1})} and even to the very purpose and the logic behind their formulation. Moreover, often, authors focus on specific subsystems \cite{virgo}, which ultimately is a way to fix physical constraints, of a kind that might be hard to translate within a formal framework.

In this paper, we will only refer to the first two acceptations, { holding the thermodynamic forces as constraints. ``Dual'' aspects of flow-driven systems \cite{kawazura2,duality} will not be considered here}. We will employ the following frameworks to describe nonequilibrium systems (from the most general to the least):
\begin{itemize}
\item[-] Phenomenological balances of fluxes and forces;
\item[-] Linear networks (there is some confusion in the literature between phenomenological principles {\it \'a la} Prigogine and network principles \cite{niven}; we comment on this in Section~\ref{maxmin}) (employing the electric circuit analogy);
\item[-] (Stochastic) Thermodynamics of Markov processes.
\end{itemize}
None of them (except the first, especially the latter) needs be \textit{the} framework for nonequilibrium phenomena. So, all propositions below have to be put into context. However, given the powerfulness of these frameworks (except the first, especially the latter), when dealing with actual systems, an effort should be made to qualitatively inscribe the process of interest within one of them. Otherwise, all conclusions are doomed to either be circumscribed within a very specific range or to be so general as to be unintelligible (cf. \cite{swenson}).

Our main theses are:
\begin{itemize}
\item[-] Ziegler's principle implies reciprocal relations between currents and forces (Sections~\ref{ziegler} and \ref{ortho}).
\item[-] Ziegler's principle does not hold beyond the linear regime (Section~\ref{ortho}).
\item[-] It is inappropriate to suppose that forces are functions of the currents only (Section~\ref{orthocon}).
\item[-] Ziegler's principle does not imply the onset of steady states (Section~\ref{kirchhoff}).
\item[-] Currents do not ``tend to'' maximize entropy production (Section~\ref{tendto}).
\end{itemize}

Also, in Section~\ref{maxmin}, we compare MaxEP and MinEP, showing that they have completely different setups. According to the former, the variable quantities are the response coefficients at fixed external forces, given the steady state; response relations follow. According to the latter, the variable quantities are the currents, at fixed external forces, given the response coefficients; the steady state follows. {
In a way, MaxEP is a constitutive principle that makes a statement about how the microphysics of the problem and the network topology affect the ``inertia'' of a system, while MinEP is a principle prescribing the behavior of states under given phenomenological laws. As discussed in Section~\ref{tendto}, evolution towards a state of minimum or maximum entropy production rate depends on what is susceptible of time variation, either the structure of the phenomenological laws or the state of the system.}

All results rely on a coarse proof of Ziegler's principle in the linear regime, exposed in the preparatory Section~\ref{ziegler}, where it is shown that any departure from Onsager's reciprocal relations decreases the entropy production, at fixed external forces. Technical Sections \ref{ortho} and \ref{kirchhoff} are independent of one another and can be safely skipped in view of our conclusions. We will employ Einstein's convention on index contraction wherever possible; $(a_1 a_2\ldots,i_1 i_2\ldots,b_1 b_2 \ldots)$ (resp.
 $[\ldots]$) denotes complete symmetrization (resp. antisymmetrization) of all the indices to the left and to the right of commas. The equivalence sign,~``$\equiv$'', is used to impose constraints. By ``stationary'' (denoted with $^\ast$), we mean extremal solutions of a variational principle, while ``steady'' (denoted with $^{(s)}$) designates a configuration of currents that obeys Kirchhoff's current law.

\section{\label{ziegler}A Proof of Ziegler's Principle}

Our first contribution is a rephrasing of Ziegler's principle in the linear regime, which is shown to provide Onsager's reciprocal relations. Whilst the result is well-known and the derivation is not particularly elegant, this perspective will prove useful in the following.

The setup is as follows. Fluxes, $J_i$, flow within a system, induced by conjugate thermodynamic forces, $F^i$. In principle, index $i$ may account for both discrete vector indices and spatial coordinates, in the case of continuous media, but we will only work with a finite number of conjugate thermodynamic variables. The entropy production, or power output, is defined by the bilinear form:
\be
\sigma = F^i J_i
\ee
(Einstein's convention on index contraction is employed). In this expression, forces and currents are independent variables. In the linear regime, dissipation is quantified by the dissipation function:
\be
\omega = L^{ij} J_i J_j \label{eq:power}
\ee
which grows quadratically with the currents, being $\bs{L}= \{L^{ij}\}_{i,j}$ a known positive definite matrix. Notice that the dissipation function only depends on the currents. We will comment on this assumption beyond the linear regime in the discussion at the end of Section~\ref{ortho}.

How will forces, $F^i = F^i(\bs{J})$, be related to currents in the linear regime? Ziegler's principle, as reformulated, {e.g.}, in \cite{martyu}, states that physical currents maximize dissipation, given that all the entropy production goes out as dissipated power:
\be
\omega(\bs{J}) \equiv F^i(\bs{J}) J_i \label{eq:powdiss}
\ee
The stationary solution is:
\be
J^\ast_i = L_{ij} F^j\label{eq:stationary}
\ee
where lowering indices denotes matrix inversion, $\bs{L}^{-1}= \{L_{ij}\}_{i,j}$. The standard proof of this result is as follows. One keeps track of the constraint, $\sigma - \omega \equiv 0$, by introducing a Lagrange multiplier, $\lambda$, and then sets the variation of $\sigma + \lambda (\sigma - \omega)$ with respect to the currents to zero, obtaining:
\be
F_i + \lambda(F_i - 2 L_{ij} J^j) = 0 \label{eq:lagrange}
\ee
whence it follows that:
\be
J^\ast_i = \frac{1 + \lambda}{2\lambda} L_{ij} F^j
\ee
The value of the Lagrange multiplier is determined by plugging the solution into the constraint equation,
yielding $\lambda = 1$ and $J^\ast_i = L_{ij} F^j$. The fact that the stationary solution is a maximum is easily seen by taking the second derivative of Equation~(\ref{eq:lagrange}) at the stationary point.

While this proof is simple and elegant, its meaning is somewhat obscure. Equation (\ref{eq:stationary}) is such an obvious choice, given $\sigma \equiv \omega$, that the result almost seems tautological. However, it is not, as the following alternative proof highlights. Let us enforce the constraint from the very beginning. The most general relation between currents and forces satisfying $\sigma \equiv \omega$ is: 
\be
F^i = M^{ij} J_j \label{eq:linres}
\ee
where $\bs{M} = \{M^{ij}\}_{i,j}$ is a generic linear response matrix, which is assumed to be invertible. Replacing this relation into $\sigma$ and imposing $\sigma \equiv \omega$, we obtain $M^{ij} J_j J_i \equiv L^{ij} J_i J_j$. It follows that the symmetric part of $\bs{M}$ coincides with $\bs{L}$; in other words:
\be
\bs{M} = \bs{L} + \bs{\Omega}
\ee
with $\bs{\Omega}= \{\Omega^{ij}\}_{i,j}$ being a skew-symmetric matrix. For example, for a system with two conjugate pairs of variables, the most general expression for the currents that is compatible with the constraint is given~by:
\begin{subequations} \label{eq:twocur}
\bea
J_1 & = & L_{11} F_1 + (L_{12} + \Omega) F_2 \\
J_2 & = & (L_{12} - \Omega) F_1 + L_{22} F_2
\ees
Notice that when $\Omega > 0$, force, $F_2$, enhances current, $J_1$, more than force, $F_1$, enhances current, $J_2$. Out of the special case, entropy production can now be expressed in terms of the variables, $\Omega^{ij}, i< j$, and the fixed forces as:
\bea
\sigma(\bs{\Omega}) = M_{ij} F^i F^j
\eea
We now let $\bs{\Omega} \to \epsilon \, \bs{\Omega} $ with $\epsilon$ be a small parameter, and let us look for variations of the entropy production to the second order in $\epsilon^2$. We need to invert $\bs{M}$:
\be
\bs{M}^{-1} = \bs{L}^{-1} - \epsilon \bs{L}^{-1} \bs{\Omega} \, \bs{L}^{-1} + \epsilon^2 \bs{L}^{-1} \bs{\Omega} \, \bs{L}^{-1} \bs{\Omega} \, \bs{L}^{-1} + O(\epsilon^3)
\ee
Due to the skew-symmetry of the first order term, the first correction to the entropy production is   second-order. Since $\bs{L}$ is a positive definite, it admits a square root, $\bs{L}^{1/2}$, and so does its inverse. With a rescaling, $\vec{f} = \bs{L}^{-1/2} \vec{F}$ and $\bs{\varepsilon} = \epsilon \bs{L}^{-1/2} \bs{\Omega} \, \bs{L}^{-1/2} $, we obtain:
\be
\sigma(\bs{\Omega}) - \sigma(\bs{0}) = \vec{f} \cdot \bs{\varepsilon}^2 \vec{f} \leq 0\label{eq:ziegler}
\ee
where the inequality follows from the fact that, being that $\bs{\varepsilon}$ is skew-symmetric, its square is \mbox{negative-definite.}

Equation (\ref{eq:ziegler}) provides a local proof of Ziegler's principle, showing that as we displace currents from their stationary value in Equation (\ref{eq:stationary}) by slightly modifying the linear response relation, at fixed external forces, entropy production decreases. Thus, Ziegler's principle enforces Onsager's reciprocal relations, namely, the fact that the linear response matrix, $\bs{M}$, is symmetric. It states that any deviation from the Onsager's linear response relations, at fixed external forces, decreases the entropy production.

\section{\label{ortho}Violation of Thermodynamic Orthogonality beyond the Linear Regime}

It is a natural question whether MaxEP holds beyond the linear regime. We discuss here this issue, using the equivalent formulation as a principle of thermodynamic orthogonality, which states that phenomenological forces are orthogonal to the hypersurfaces of equi-entropy production \cite{ziegler}. It has been speculated \cite{martyu,fatigue} that orthogonality might be a general guiding principle to obtain response relations between forces and currents.
In this section, we {discuss a simple special example drawn from turbulent fluid mechanics where pushing Ziegler's principle beyond the linear regime yields inconsistencies, and generalizing, we show that it} entails a cornucopia of rather stringent reciprocal relations for higher-order response coefficients, when the forces are assumed to be analytic in the currents at $\bs{J}=0$. We argue that such relations might lead to violations of positivity of the entropy production. We then consider a particular instance of detailed-balanced Markovian dynamics on a discrete space with three states whose thermodynamics are described in the context of Schnakenberg's theory \cite{schnak}, showing how, already, for this very simple system, higher-order reciprocal relations do not hold.

\subsection{Thermodynamic Orthogonality}

Ziegler's principle of thermodynamic orthogonality \cite{ziegler} asserts that nonequilibrium thermodynamic forces, $F^i$, are orthogonal to the iso-hypersurfaces, where $\sigma$ has constant value:
\be
F^i = \lambda\frac{\delta \sigma}{\delta J_i} \label{eq:ortho}
\ee
The scalar, $\lambda(\bs{J})$, generally depends on the currents. Maximizing entropy production while holding Equation~(\ref{eq:powdiss}) as a constraint via a Lagrange multiplier, $\mu$:
\be
\frac{\delta }{\delta J_k} \left\{ F^i J_i + \mu [ \sigma(\bs{J}) - F^i J_i] \right\} = 0
\ee
one obtains Equation~(\ref{eq:ortho}), with $\lambda = \mu/(\mu-1)$ to be determined by replacing the solution into the constraint. Reference \cite{martyu} discusses at length the nature of the maximum of the extremal. Hence, thermodynamic orthogonality follows from MaxEP.

{
That Ziegler's principle has problems beyond the linear regime can already be seen with the following example. Consider two independent fluid flows in pipes, respectively, in laminar and fully developed turbulent regimes, in which case the thermodynamic force is (proportional to) the pressure loss \cite{niven}. The first case corresponds to the linear regime, while in the second, the pressure loss is proportional to $J_t|J_t|$. Let $J_{t} >0$ for definitiveness. Assuming that the two systems are completely unrelated among themselves, the total entropy production rate will be additive:
\be
\sigma = M_l J_l^2 + M_t J_t^3
\ee

Applying Ziegler's principle yields for the MaxEP forces:
\be
F_l = 2\lambda(\bs{J}) M_l J_l, \quad F_t = 3\lambda(\bs{J}) M_t J_t^2
\ee

It is clear that there is no analytical function of the currents that makes $\sigma = F_l J_l + F_t J_t$, unless either $M_l = 0$ or $M_t = 0$. In particular, the MaxEP forces are not analytical in the currents. However, this is in disagreement with the fact that since the two systems are uncorrelated, the two thermodynamic forces are known to coincide with the pressure losses, which are respectively linear and quadratic. In general, while MinEP and MaxEP do work when the entropy production is a homogeneous polynomial in the currents, they do not in all other cases, as commented in \cite{niven}.}

\subsection{Higher-Order Reciprocal Relations}

Let us now assume that the forces always admit a Taylor expansion at $\bs{J} = 0$:
\be
F^a = \sum_{r \geq 2} L^{a i_2 \ldots i_r} J_{i_2} \ldots J_{i_r}\label{eq:forces}
\ee
Here, $L^{i_1 \ldots i_r}$ is a rank-$r$ tensor, symmetric in its last $r-1$ indices (except the first) ({{Notice that in the case of continuous systems, when there are no other preferred reference vector fields, one will have it that the direction of the force field is determined by that of the current field and that the response coefficient will only depend
on the modulus of $\vec{J}$, so that only even contributions to the Taylor expansion will appear. This is not the case in the discrete case (as in the hydrodynamic example above and in the master equation example that we will discuss below).}}). The entropy production now reads:
\be
\sigma = \sum_{r \geq 2} M^{i_1 i_2 \ldots i_r} J_{i_1} \ldots J_{i_r} \label{eq:ep}
\ee
and its derivatives:
\be
\frac{\delta\sigma}{\delta J_a} = \sum_{r \geq 2} r M^{a i_2 \ldots i_r} J_{i_2} \ldots J_{i_r}
\ee
where the totally symmetric $r$-tensors appearing in the right-hand side are obtained by symmetrizing the response tensors with respect to the first index:
\be
M^{a i_2 \ldots i_r} = r^{-1} \left[ L^{a i_2 \ldots i_r} + (r-1) L^{i_2 \ldots i_{r} a} \right]
\ee
For the orthogonality relation Equation~(\ref{eq:ortho}) to be satisfied, there must exist a function, $\lambda(\bs{J})$, such that:
\be
\lambda = \frac{F^{a}}{\delta \sigma / \delta J_{a}} = \frac{\sum_{r \geq 2} L^{a i_2 \ldots i_r} J_{i_2} \ldots J_{i_r}}{ \sum_{r' \geq 2} r' M^{a i'_2 \ldots i'_r} J_{i'_2} \ldots J_{i'_r}}
\ee
for all $a$ (here, index $a$ is mute, that is, it is not summed over). Thus, for any two indices, $a$ and $b$, the~conditions:
\be
F^{a} \frac{\delta \sigma }{\delta J_{b}} - F^{b} \frac{\delta \sigma}{\delta J_{a}} = 0
\ee
must be satisfied, order-by-order, in the currents. At order $s$, one finds:
\be
T^{abi_1\ldots i_s} := \sum_{t = 0}^{s-2} (t+2) L^{[a, (i_1 \ldots i_{s-t-1},} M^{,b], i_{s-t} \ldots i_{s})} = 0
\ee

These are a set of higher-order reciprocal relations.
At the second order:
\be
T^{abi_1i_2} = L^{a i_1} L^{i_2b} + L^{a i_2} L^{i_1b} - L^{i_2 a} L^{bi_1} - L^{i_1 a} L^{b i_2} = 0 \label{eq:second}
\ee
Considering $(i_1,i_2) = (a,b)$, one finds $L^{ab} = \pm L^{ba}$, and considering only $i_1 = a$, one retrieves Onsager's relations. \textit{Vice versa}, given Onsager's relations, Equation~(\ref{eq:second}) is satisfied. At the third order,
\be
2 L^{[a,(i_1 i_2,} M^{,b],i_3)} + 3 L^{[a,(i_1,}M^{,b],i_2 i_3)} = 0
\ee
Notice that third-order coefficients are coupled to second-order ones, and so on, at higher orders. Let us write down explicitly these expressions for a system with two conjugate forces $(F^1,F^2)$ and   currents $(J_1,J_2)$, yielding four independent relations:
\bes
L^{111} L^{12} & = & L^{11} (2 L^{121} - L^{211}) \\
L^{222} L^{12} & = & L^{22} (2 L^{212} - L^{122}) \\
L^{111} L^{22} & = & L^{12} (2 L^{112} - 3 L^{211}) + 2 L^{11} L^{122} \\
L^{222} L^{11} & = & L^{12} (2 L^{221} - 3 L^{122}) + 2 L^{22} L^{211}
\ees

To appreciate the backreaction of higher-order response relations onto lower-order ones, notice that, if $L^{111} = L^{222} = 0$ (as will be the case in the example below), employing the above relations, one obtains:
\be
L^{11} L^{22} - (L^{12})^2 = 0
\ee
where we recognize in the left-hand side the determinant of the linear response matrix, which is then necessarily degenerate. Along the direction where the null eigenvector of the linear response matrix lies, the entropy production is a cubic function near the origin, and it can then be made negative, unphysically.

\subsection{Counterexample}

We will now give a simple example where higher-order reciprocity relations do not hold, in the context of the ensemble thermodynamics of master equation systems \cite{esposito}, whose stochastic counterpart along individual jump trajectories is more and more becoming {\it the} canonical formulation of nonequilibrium thermodynamics \cite{seifert}. For simplicity, we consider a system with uniform transition rates, $w_{yx} = 1$, between any two vertices, $x,y$, in a graph, in which case, the dynamics satisfy detailed balance and affords the uniform (unnormalized) equilibrium steady state, $p^{(s)}_x = 1$. We perturb the state of the system out of the steady state, $p_x = 1 + \epsilon_x$, where $\sum_x \epsilon_x = 0$. Currents and forces along the edges of the graph are widely agreed to be defined according to Schnakenberg's theory \cite{schnak} as:
\begin{align}
j_{yx}  ~=~ & p_x - p_{y} ~=~ \epsilon_x - \epsilon_{y} \label{eq:currs} \\
f_{yx}  ~=~ & \log \frac{p_x}{p_{y}} ~\approx~ j_{yx} \left[1 - (\epsilon_x + \epsilon_{y})/2 
\right]
\end{align}

In the second expression, we expanded up to the second order in $\epsilon$, given that the currents are themselves the first order. Let us now focus on a three-state system, $x = 1,2,3$. Notice that when the state of the system is perturbed, keeping the rates fixed, Kirchhoff's loop law (KLL) for the forces holds along the cycle:
\be
f_{12} + f_{23} + f_{31} = 0
\ee
so that only two forces are independent. We choose $F^1 = f_{21} = -f_{12}$ and $F^3 = f_{23}$. The entropy production reads:
\be
\sigma = f_{12} j_{12} + f_{23} j_{23} + f_{31} j_{31} = F^1 J_1 + F^3 J_3
\ee
where we identified the conjugate currents, $J_1 = j_{21} + j_{31}$ and $J_3 = j_{23} - j_{31}$. Reference \cite{duality} discusses some theoretical reasons why these observables are fundamental when displacing from the equilibrium steady state. We will comment below on the generality of this choice.

Since also the currents obey KLL, $j_{12} + j_{23} + j_{31} = 0$, we obtain:
\bea
J_1 & = & 2 j_{21} - j_{23} \\
J_3 & = & 2 j_{23} - j_{21}
\eea
Similarly, using Equation~(\ref{eq:currs}) and $\epsilon_1 + \epsilon_2 + \epsilon_3 = 0$, we can express the population increments in terms of the fundamental currents:
\bea
\epsilon_1 & = & (-j_{23} + 2 j_{21})/3 = J_1/3 \\
\epsilon_3 & = & (-j_{21} + 2 j_{23})/3 = J_3/3
\eea
yielding the response relations: 
\bes
F^1 & = & \frac{2 J_1 + J_3}{3} \left( 1 + \frac{J_3}{6} 
\right) \\ 
F^3 & = & \frac{2 J_3 + J_1}{3} \left( 1 + \frac{J_1}{6} 
\right) 
\ees

Let us write explicitly the entries or the first-order response tensor:
\bea
L^{11} = \tfrac{2}{3}, \qquad L^{13} = L^{31} = \tfrac{1}{3}, \qquad L^{33} = \tfrac{2}{3}
\eea
Onsager's relations hold. The entries of the second-order tensor are:
\bes
& & L^{111} = 0, \qquad L^{113} = \tfrac{1}{18}, \qquad L^{133} = \tfrac{1}{18} ,	\\
& & L^{333} = 0, \qquad L^{331} = \tfrac{1}{18}, \qquad L^{311} = \tfrac{1}{18}.
\ees
We are precisely in the conditions where third-order relationships would dictate a vanishing determinant for the linear response matrix, which is not the case. Hence, the example violates the reciprocal relations, the orthogonality principle and MaxEP.

\subsection{\label{orthocon}Discussion}

One might dispute that failure of higher-order reciprocal relations in our simple model is due to the choice of conjugate observables. Although, this is not the case: any other choice leads to the same violation. In fact, the most general transformation of conjugate variables that leaves the entropy production invariant is an invertible linear transformation, $F^{i'} = \Lambda^{i'}_i F^i$ and $J_{i'} = \Lambda_{i'}^{i} J_i$, with $\Lambda^j_{i'} \Lambda_i^{i'} = \delta_{i}^j$. By simple manipulations, we obtain:
\be
T^{a'b'i'_1\ldots i'_s} =
\Lambda^{a'}_a \Lambda^{b'}_b \Lambda^{i_1'}_{i_1} \ldots \Lambda^{i_s'}_{i_s} T^{abi_1\ldots i_s}
\ee
In other words, the symmetric and the antisymmetric parts of tensors are irreducible (do not get mixed and cannot be made to vanish after a linear transformation). It follows that reciprocal relations hold for one set of pairs of conjugate variables if and only if they do so for any other set of linearly related pairs that leave the entropy production unchanged.

Given Equations~(\ref{eq:forces}) and (\ref{eq:ep}), one might hypothesize that the correct phenomenological relations~are:
\be
L^{a i_2 \ldots i_r} \stackrel{?}{=} M^{a i_2 \ldots i_r} \label{eq:hypo}
\ee
or, in other words, that the response coefficients are completely symmetric with respect to all indices, including the first. This holds true for a set of third-order coefficients in the above example, given, in particular, the relation $L^{113}=L^{311}$. Although, this is an artifact of the symmetry of the problem. We report, without giving details, that slight generalizations of the above example (for example, by   setting $w_{12} = w_{12} = 2$) already invalidate this hypothesis.

Let us digress on the assumption ``all power is dissipated''. Its ultimate meaning is that there always exists a function of the currents that describes thermodynamics out of the steady state, hence that, in principle, forces can always be expressed in terms of currents. In the context of   Stochastic Thermodynamics, this is indeed the case in all situations where one displaces the system out of the steady state, yet maintaining transition rates. However, for more general situations, there is growing evidence that currents alone are not sufficient to characterize nonequilibrium states and that one should generalize to $F^a = F^a(\bs{J},\bs{K})$, where $\bs{K}$ is a set of fluctuating variables that are symmetric under time reversal \cite{zia,maes}.

Finally, we point out that there are special cases where the orthogonality principle trivially applies: when the entropy production is a homogeneous polynomial of degree, $r$, in the currents; then only one pair of conjugate force and current is considered. {Besides these special cases, we do not know of systems for which the generalized reciprocal relations hold, and we would presume that, if there exists any, they would be very rare, given that these relations are extremely restrictive.}

\section{\label{kirchhoff}Ziegler's Principle and Kirchhoff's Laws}

The kind of proposition that we address in this section is ``the steady state of open thermodynamic systems with sufficient degrees of freedom are maintained in a state at which the production of entropy is maximized given the constraints of the system'' \cite{kleidonent}.

Linear networks are often employed in irreversible thermodynamics as archetypical models, where steady states are realized in the form of conservation laws at the nodes. Although they regard the flow of the same quantity (charge, mass, \emph{etc}.) along a system of conduits rather than fluxes of several species, they display the full phenomenology of more general nonequilibrium systems. In this context, \v Zupanovi\'c and coworkers \cite{looplaw} asserted that Kirchhoff's loop law (KLL) follows from MaxEP, admitting as a further constraint Kirchhoff's current law (KCL). This result is also reported by Bruers \cite{bruers2}. The proof is deceptively correct, and it leads to the conclusion that MaxEP is relevant to nonequilibrium steady states. However, there are some subtle issues concerning their derivation. It follows from our analysis that Ziegler's principle and Kirchhoff's Laws are independent facts. What happens when imposing KCL is that a set of coarse-grained observables is selected. They are defined along cycles of the network, and they pertain to steady states. MaxEP then implies Onsager's relations among such macroscopic observables, but it cannot be held responsible for the onset of the steady state---that is due to MinEP.

\subsection{Reciprocal Relations for Loop Currents and Forces}

In a linear network microscopic currents, $j_e$, flow along the edges, $e$, of a graph; we collect them in a vector, $\bs{j} $. Conjugate to them are microscopic forces, $\bs{f}$. The entropy production is given by $\sigma = \bs{f} \cdot \bs{j}$ and the dissipation function by $\omega = \bs{j} \cdot \bs{\ell} \bs{j}$. Usually, the symmetric bilinear form, $\bs{\ell}$, is diagonal (this is the case in resistor networks and for Markov chains), but we can slightly generalize in the present context. At a steady state, currents obey conservation laws at the vertices, $x$, of the network, according to KCL:
\be
\bs{\partial} \bs{j}^{(s)} = 0
\ee
where $\bs{\partial} = (\partial_x^{\, e})_{x,e}$ is the incidence matrix of the graph ($-1$ if edge, $e$, goes into vertex, $x$, $+1$ if it comes out of it, and zero, otherwise). According to Schnakenberg's theory \cite{schnak}, letting $X$ be the number of vertices and $E$ the number of edges, KCL expresses $X-1$ independent constraints that allow the casting of the steady current vector as a linear combination of $E-X+1$ macroscopic currents, $J_i$, which characterize the steady state:
\be
\bs{j}^{(s)} = J_i \, \bs{c}^i\label{eq:KCL}
\ee
Here, $\bs{c}^i = (c^i_e)_e$ are $E-X+1$ independent vectors in the kernel of the incidence matrix; their graphical interpretation is as cycles. Here, we rely on a purely algebraic treatment; see \cite{myself} for a concise and complete discussion of the graphical construction, which, by the way, is not strictly necessary here.

Kirchhoff's loop law (KLL) states that the fixed, but arbitrary, macroscopic forces conjugate to the macroscopic currents, $J_i$, can be expressed as sums over cycles of the local forces:
\be
F^i = \bs{c}^i\cdot \bs{f}
\ee
For example, in an electric network, the sum of the voltage drops at the ends of resistors along a cycle (on the right) equals the electromotive force (on the left).

We now go through Ziegler's MaxEP principle by plugging constraints by hand, the constraints being: (i) KCL, namely Equation~(\ref{eq:KCL}); and (ii) ``all power dissipated'', $\omega - \sigma \equiv 0$. Imposing constraint (i) on the dissipation function and the entropy production, by simple algebraic manipulations, we find:
\be
\sigma = F^i J_i, \qquad \omega = L^{ij} J_i J_j \label{eq:entconst}
\ee
where:
\be
L^{ij} = \bs{c}^i \cdot \bs{\ell} \bs{c}^{j}
\ee
In Equation~(\ref{eq:entconst}), KCL is used to express both entropy production and dissipated power in terms of macroscopic forces and currents. Notice that we have not maximized yet, but macroscopic forces already loom as variables conjugate to the macroscopic currents, according to KLL. According to our discussion in Section~\ref{ziegler}, imposing constraint (ii), we find that $F^i = M^{ij} J_j$, where the symmetric part of $M^{ij}$ coincides with $L^{ij}$. Now, we are ready to maximize entropy production. It follows from MaxEP that the skew-symmetric part of $ M^{ij}$ vanishes. In other words, Onsager's relations for the linear response coefficients between macroscopic currents and forces hold.

We cannot come to the same conclusion as regards microscopic linear response relations. Given $\bs{f} = \bs{m} \bs{j}$, with the diagonal part, $(m^{ee'} + m^{e'e})/2 = \ell^{ee'}$, and a nonvanishing off-diagonal part, we~obtain:
\be
F^i = \bs{c}^i \cdot \bs{f}
= \Big( L^{ii'} + \sum_{e \neq e'} \frac{m^{ee'} - m^{e'e}}{2} c_e^i c_{e'}^{i'} \Big) J_{i'}
\ee
The second term in the last passage vanishes; hence, the above relation does not force the skew-symmetric part of the microscopic linear response coefficients to vanish. In other words, by imposing KCL, we have coarse-grained the description from microscopic to macroscopic currents and~forces.

{ \subsection{\label{maxmin}Discussion: MaxEP \emph{vs}. MinEP, Phenomenological \emph{vs}. Network Principles.}}

We have proven that, implementing the current conservation law at the nodes of a network, MaxEP prescribes reciprocal relations among a coarse-grained set of observables, without specifying microscopic ({\it i.e.}, edge-by-edge) reciprocity relations. Notice that the maximization procedure {\it per se} is not responsible for the onset of the steady state.

In \cite{myself}, it is proven that the steady state (KCL) follows from MinEP, given constrained values of the macroscopic forces (KLL); in that case, the constraint participates in the minimization procedure in a crucial way. Let us briefly review the result. For a given set of local linear response coefficients, $\bs{m} = \bs{\ell}$, we minimize the entropy production holding fixed values of the macroscopic forces:
$\bs{c}^i \cdot \bs{f} \equiv F^i$. Setting
\be
\frac{\delta}{\delta \bs{j}} \left( \bs{j} \cdot \bs{\ell} \bs{j} + \lambda_i \, \bs{c}^i \cdot \bs{\ell} \bs{j} \right) = 0
\ee
we obtain $\bs{j}^\ast = - \lambda_i \, \bs{c}^i /2$. We recognize KCL as stated in Equation~(\ref{eq:KCL}), given $\lambda_i = -2 J_i$.

In light of these results, let us analyze the proposition ``currents in network branches are distributed in such a way as to achieve the state of maximum entropy production'' \cite{looplaw}, which conveys that a departure from the steady state ({\it i.e.}, breaking KCL) will diminish the entropy production. This is not the case; rather, departing from the steady state, yet keeping the macroscopic external forces fixed, entropy production increases. MaxEP in the presence of KCL states that, maintaining the steady state, if reciprocal relations are slightly violated at fixed external forces, then entropy production decreases.

 {With reference to electric networks, the physical picture behind MinEP is the following: one prepares the system in a state where charges are concentrated at the nodes, and then lets them free to adjust, so as to cancel the electric field inside the conductor. Holding the fixed electromotive forces as constraints, entropy production is minimized to its steady value along this process. Notice that, differing from MaxEP, the topological properties of the network and Ohm's relation (which follows from Maxwell's equations) are held fixed.}

{

An important issue emerges as to what is the relationship between network-based and   phenomenological principles. Niven (\cite{niven}[Section~2]) was particularly careful in warning about the distinction between different frameworks. While network thermodynamics deals with one physical species being conserved at the nodes of a network, phenomenological principles regard different species (mass in incompressible fluid dynamics, charge in electrodynamics \emph{etc}.). In a paper by the author~(\cite{myself}[Section~IV.a]), it is shown that, according to a graph-theoretical construction, whose details are given in \cite{duality}, network currents and forces can be rearranged, as in Prigogine's phenomenological formulation of MinEP~(\cite{prigo}[\S VI.2.]), so that network thermodynamics can be seen as a special case of the more generic phenomenological thermodynamics. {\it Vice versa}, can phenomenological thermodynamics always be put in the form of network thermodynamics? To the author's knowledge, there is yet no such general principle, but there are hints that this equivalence can be achieved. The intuitive picture is as follows. Master-equation thermodynamics involves generic probability flows on an abstract state space. Labeling state $i = (E,N,Q,...)$ to account for energy, number of particles, electric charges \emph{etc}., the marginal averages of the probability current will yield phenomenological currents. According to the paradigm of {\it local detailed balance} \cite{esposito}, the entropy production accounts for currents of different species. The case could be made that Prigogine's phenomenological MinEP might descend from network MinEP when a physical texture is bestowed on probability currents.}

\section{\label{tendto}Conclusions: Currents do not ``Tend to'' Maximize Entropy Production}

Many an author have speculated that Ziegler's principle corroborates that currents {\it arrange themselves}, so as to achieve the state of maximum entropy production \cite{looplaw} or \textit{tend to} to the state with maximum entropy along the shortest possible path \cite{martyu}, bestowing a dynamic nature to what in Section~\ref{ziegler} was formulated as a purely static principle. This eventually leads to a clash with MinEP, which states that under suitable hypothesis on the dynamics, currents tend to minimize the entropy production.

It is then natural to ask, in the evolution of a physical system, which of the two occurs: do response coefficients relax at fixed macroscopic forces, at fixed conservation laws, maximizing entropy production, or do currents relax, at fixed external macroscopic forces, given fixed response coefficients?

In this respect, physical intuition may lead one astray: one might, in fact, oppose that, if forces and response relations are fixed, then currents are fixed; and that the system relaxes by modifying the relationship between forces and currents. However, response relationships between currents and forces are due to the microphysics of the problem, for example, the material properties of an electric network, and should not depend on the state of the system (resistances are given in the time dynamics of electric networks). This apparent paradox is easily solved by considering that only {\it some} of the forces are fixed (in an electric network, only the electromotive forces, not all voltage drops). Initial conditions out of the steady state imply that a set of ``internal'' forces will contribute to the total entropy production out of the steady state.

A framework where this mechanism is realized is that of Markovian processes on networks. The author's analysis in (\cite{myself}[III.B]) shows that such systems have fixed linear response relationships, a fixed set of external macroscopic forces and the system tends to minimize entropy production as currents relax to their steady value. Nevertheless, one might envisage a framework where response coefficients (encoding the microphysics of the system) vary in time at fixed ``external'' forces. In that case, one would indeed see a dynamical realization of MaxEP. The situation is one in which Onsager's relations are violated at some initial time, and then the system arranges itself in such a way as to restore them in the long time limit.

In a way, our interpretation makes precise an intuition of Kleidon and Dyke, who commented: ``While we acknowledge that there is still some way to go before the Maximum Entropy Production principle is firmly established on analytical grounds, we believe that seeking a formulation of the principle that will guarantee the production of accurate predictions for the steady states of real world systems to be fundamentally misguided'' \cite{kleidon}.

As a final remark, we point out that several controversial topics that were not addressed are: Dewar's formulation of MaxEP, its nature as an inference method and applications of MaxEP to climate and ecological models. About the former, we point out a rich bibliography on the Maximum Caliber principle~\cite{marzen,wu,stock}, where it is shown that the Markovian path measure indeed maximizes path entropy for given constraints, but in that case, path entropy has nothing to do with entropy production, so one cannot infer MaxEP. About the latter, let us briefly linger on the very interesting system-boundary problem~\cite{virgo}. It is indeed plausible that drawing suitable boundaries within the system might result in an effective description, where a local measure of entropy production is maximized at the steady state. We report that in the context of linear network thermodynamics and of stochastic thermodynamics, the author has long tried to abstractly draw boundaries between an effective environment and an effective system, defining an effective entropy production and imposing constraints in such a way that any deviation from the physical steady currents would augment the entropy production, with no success, so far.

\section{Acknowledgments}

The research was supported by the National Research Fund Luxembourg in the frame of project FNR/A11/02.

\bibliographystyle{mdpi}

\begin{thebibliography}{----}

\bibitem{ziegler} Ziegler, H. {\it An Introduction to Thermomechanics}; North-Holland: Amsterdam, The Netherlands, 1983.

\bibitem{ozawa} Ozawa, H.; Ohmura, A.; Lorenz, R.; Pujol, T. The second law of thermodynamics and the global climate system: A review of the maximum entropy production principle. {\it Rev. Geophys.} {\bf 2003}, {\it 41}, doi:10.1029/2002RG000113.

\bibitem{kleidon} Dyke, J.; Kleidon, A. The maximum entropy production principle: Its theoretical
foundations and applications to the earth system. {\it Entropy} {\bf 2010}, {\it 12}, 613--630.

\bibitem{popper} Popper, K. {\it Conjectures and Refutations: The Growth of Scientific Knowledge}; Routledge: London, UK, 1963.

\bibitem{myself} Polettini, M. Macroscopic constraints for the minimum entropy production principle. {\em Phys. Rev. E} {\bf 2011}, {\em 84}, 051117.

\bibitem{bruers2} Bruers, S.; Maes, C.; Neto\v cn\'y, K. On the validity of entropy production principles for linear electrical circuits. {\it J. Stat. Phys.} {\bf 2007}, {\it 129}, 725--740.

\bibitem{luo} Jiu-Li, L.; van den Broeck, C.; Nicolis, G. Stability criteria and fluctuations around nonequilibrium states. {\it Zeitschrift f\"ur Physik B Condensed Matter} {\bf 1984}, {\it 56}, 165--170.

\bibitem{ma} Maes, C.; Neto\v cn\'y, K. Minimum entropy production principle from a dynamical fluctuation law. {\it   J. Math. Phys.} {\bf 2007}, {\it 48}, 053306.

\bibitem{martyu} Martyushev, L.M.; Seleznev, V.D. Maximum entropy production principle in physics, chemistry and biology. {\em Phys. Rep.} {\bf 2006}, {\em 426}, 1--45.

\bibitem{looplaw} \v Zupanovi\'c, P.; Jureti\'c, D.; Botri\'c, S. Kirchhoff's loop law and the maximum entropy production principle. {\em Phys. Rev. E} {\bf 2004}, {\em 70}, 056108.

\bibitem{dewar} Dewar, R.C. Maximum entropy production and the fluctuation theorem. {\em J. Phys. A Math. Gen. } {\bf 2005}, {\em 38}, doi:10.1088/0305-4470/38/21/L01.

\bibitem{swenson} Swenson, R. Autocatakinetics, evolution, and the law of maximum entropy production: A principled foundation towards the study of human ecology. {\it Adv. Human Ecol.} {\bf 1997}, {\it 6}, 1--48.

\bibitem{zupa} \v Zupanovi\'c, P.; Botri\'c, S.; Jureti\'c, D. Relaxation process, MaxEnt formalism and Einstein's formula for the probability of fluctuations. {\em Croatica chemica acta} {\bf 2006}, {\em 79}, 335--338.

\bibitem{dewar2} Dewar, R.C. Maximum entropy production as an inference algorithm that
translates physical assumptions into macroscopic predictions: Don't shoot the messenger. {\it Entropy} {\bf 2009}, {\it 11}, 931--944.

\bibitem{kawazura1} Kawazura, Y.; Yoshida, Z. Comparison of entropy production rates in two different types of   self-organized flows: B\'enar convection and zonal flow. {\it Phys. Plasmas} {\bf 2012}, {\it 19}, 012305.

\bibitem{virgo} Virgo, N. From maximum entropy to maximum entropy production: A new approach. {\em Entropy} {\bf 2010}, {\em 12}, 107--126.

\bibitem{kawazura2} Kawazura, Y.; Yoshida, Z. Entropy production rate in a flux-driven self-organizing system. {\it  Phys. Rev. E} {\bf 2012}, {\it 82}, 066403.

\bibitem{duality} Polettini, M. System/environment duality of nonequilibrium network observables. {\bf 2011}, arXiv:1106.1280.

\bibitem{niven} Niven, R.K. Simultaneous extrema in the entropy production for steady-state fluid flow in parallel pipes. {\it J. Non-Equilib. Thermodyn.} {\bf 2010}, {\it 35}, 347--378.

\bibitem{fatigue} Khonsari, M.M.; Amiri, M. {\it Introduction to Thermodynamics of Mechanical Fatigue}; CRC Press: Boca Raton, FL, USA, 2012.

\bibitem{schnak} Schnakenberg, J. Network theory of microscopic and macroscopic behavior of master equation systems. {\em Rev. Mod. Phys.} {\bf 1976}, {\em 48}, 571--585.

\bibitem{esposito} Esposito, M.; van den Broeck, C. The three faces of the second law:
I. Master equation systems. {\em Phys. Rev. E} {\bf 2010}, {\em 82}, 011143.

\bibitem{seifert} Seifert, U. Stochastic thermodynamics, fluctuation theorems,
and molecular machines.
{\em Rep. Prog. Phys.} {\bf 2012}, {\em 75},  doi:10.1088/0034-4885/75/12/126001.

\bibitem{zia} Zia, R.K.P.; Schmittmann, B. A possible classification of nonequilibrium steady states. {\em J. Phys. A Math. Gen.} {\bf 2006}, {\em 39}, L407.

\bibitem{maes} Maes, C.; Neto\v cn\'y, K.; Wynants, B. On and beyond entropy production: The case of Markov jump processes. {\em Markov Process. Relat. Field.} {\bf 2008}, {\em 14}, 445--464.

\bibitem{kleidonent} Dyke, J; Kleidon, A.  The maximum entropy production principle: Its theoretical foundations and applications to the Earth System. {\it Entropy} {\bf 2010}, {\it 12}, 613--630.

\bibitem{prigo} Prigogine, I. \textit{Introduction to Thermodynamics of Irreversible Processes}; John Wiley and Sons: New~York, NY, USA, 1955.

\bibitem{marzen} Marzen, S.; Wu, D.; Inamdar, M.; Phillips, R. An equivalence between a Maximum Caliber analysis of two-state kinetics and the Ising model. {\bf 2010}, arXiv:1008.2726 [physics.bio-ph].

\bibitem{wu} Wu, D.; Ghosh, K.; Inamdar, M.; Lee, H.J.; Fraser, S., Dill, K.; Phillips, R. Trajectory approach to two-state kinetics of single particles on sculpted energy landscapes. {\em Phys. Rev. Lett.} {\bf 2009}, {\em 103},~050603.

\bibitem{stock} Stock, G.; Ghosh, K.; Dill, K.A. Maximum caliber: A variational approach applied to two-state dynamics. {\em J. Chem. Phys.} {\bf 2008}, {\em 128}, 194102--194113.

\bibitem{bruers1}
Bruers, S. A discussion on maximum entropy production and information theory. {\em J. Phys. A Math. Theor.} {\bf 2007}, {\em 40}, doi:10.1088/1751-8113/40/27/003.



\end{thebibliography}
\makeatletter
\renewcommand\@biblabel[1]{#1. }
\makeatother

\end{document}